\input{aipcheck}

\documentclass[
    ,final            
  ]
  {aipproc}
  
\usepackage{times,mathptm}

\layoutstyle{6x9}

\newcommand {\nc} {\newcommand}
\nc {\beq} {\begin{eqnarray}}
\nc {\eeq} {\end{eqnarray}}

\begin{document}

\title{
Multi-Quarks and Two-Baryon Interaction\\
in Lattice QCD}

\classification{12.38.Gc,12.38.Aw,14.20.Jn,12.39.Pn}
\keywords      {lattice QCD, multi-quarks, confinement, inter-quark potential}

\author{F. Okiharu}{
address={
Nihon University, 
1-8-14 Kanda Surugadai, Chiyoda, Tokyo 101-8308, Japan}
}

\author{H. Suganuma}{
address={
Dept. of Phys., Kyoto University, 
Kitashirakawa, Sakyo, Kyoto 606-8502, Japan}
}

\author{T. T. Takahashi}{
address={
YITP, Kyoto University, 
Kitashirakawa, Sakyo, Kyoto 606-8502, Japan}
}

\author{T. Doi}{
address={
RIKEN-BNL Research Center, BNL, Upton, New York 11973, USA}
}

\begin{abstract}
We study multi-quark (3Q,4Q,5Q) systems in lattice QCD. 
We perform the detailed studies of multi-quark potentials in lattice QCD 
to clarify the inter-quark interaction in multi-quark systems. 
We find that all the multi-quark potentials are well described 
by the OGE Coulomb plus multi-Y-type linear potential, i.e., 
the multi-Y Ansatz. 
For multi-quark systems, we observe lattice QCD evidences of ``flip-flop'', 
i.e., flux-tube recombination.
These lattice QCD studies give an important bridge between 
elementary particle physics and nuclear physics.
\end{abstract}
\maketitle

\section{Introduction}

The inter-quark force \cite{TS,OST05,R05} 
is one of the most important quantities in hadron physics 
to understand hadron properties at the quark-gluon level,  
and also plays an essential role to reveal properties of multi-quarks \cite{Belle,BABAR,Penta}. 
Nevertheless, before our studies, nobody knew the exact form of 
the confinement force in the multi-quark systems directly from QCD. 
In fact, some hypothetical forms of the inter-quark potential 
have been used in all the quark model calculations.
Then, the lattice QCD study of the inter-quark interaction 
is quite desired, and it leads to 
the proper Hamiltonian and 
a guideline to construct the QCD-based quark model for multi-quark systems. 

We show in Fig.1 our global strategy 
to understand hadron properties from QCD. 
One way is the direct lattice QCD calculations 
for low-lying hadron masses \cite{I05} and simple hadron matrix elements, 
although the wave function cannot be obtained in the path-integral formalism 
and the practically calculable quantities are severely limited.
The other way is to construct the quark model from QCD.
From the analysis of the inter-quark forces in lattice QCD, 
we extract the quark-model Hamiltonian.
Through the quark model calculation, one can obtain 
the quark wave-function of hadrons and more complicated properties of hadrons.
\begin{figure}[h]
\vspace{-0.5cm}
\centering
\includegraphics[height=4cm,clip]{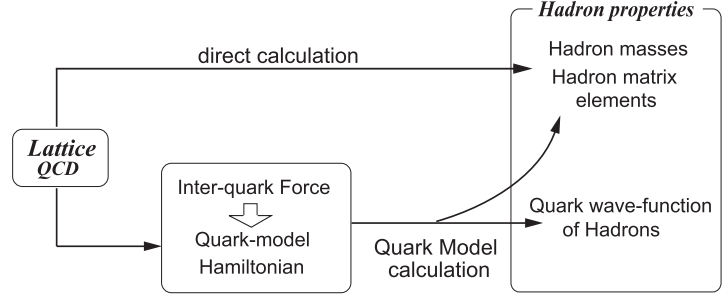}
\caption{
Our global strategy to understand the hadron properties from QCD.
}
\vspace{-0.5cm}
\end{figure}

\section{Multi-Quark Interaction from Lattice QCD}

In order to clarify the inter-quark force in multi-quark systems, 
we study the static multi-quark potentials systematically in lattice QCD 
using the multi-quark Wilson loops \cite{TS,OST05}.
As the results, 
we clarify that the multi-quark potential is well described as 
\begin{eqnarray}
V=\frac{g^2}{4\pi}\sum_{i<j}\frac{T^a_iT^b_j}{|{\bf r}_i-{\bf r}_j|}
+\sigma L_{\rm min}+C, 
\end{eqnarray}
where $L_{\rm min}$ is the minimal value of the total length of flux-tubes linking 
the static quarks. 

From lattice QCD, the Q$\bar{\rm Q}$ potential $V_{\rm Q\bar Q}$ 
is known to be well described by this form \cite{TS,R05}.
For the 3Q potential $V_{\rm 3Q}$, 
our accurate calculations clarify  
that $V_{\rm 3Q}$ is well described 
by the OGE Coulomb plus Y-type linear potential,
i.e., the Y-Ansatz\cite{TS,OST05}, which 
supports the Y-shaped flux-tube formation \cite{TS,OST05,IBSS03}. 
Note that the Y-type linear potential implies 
existence of the three-body interaction. 
(In usual many-body systems, the main interaction 
is described by two-body interactions and 
the three-body interaction is a higher-order contribution. 
In contrast, as is clarified by lattice QCD study, 
the quark confinement force in baryons is a genuine three-body interaction.)

\begin{figure}[h]
\vspace{-0.5cm}
\begin{minipage}{50mm}
\centering
\includegraphics[height=3.8cm,clip]{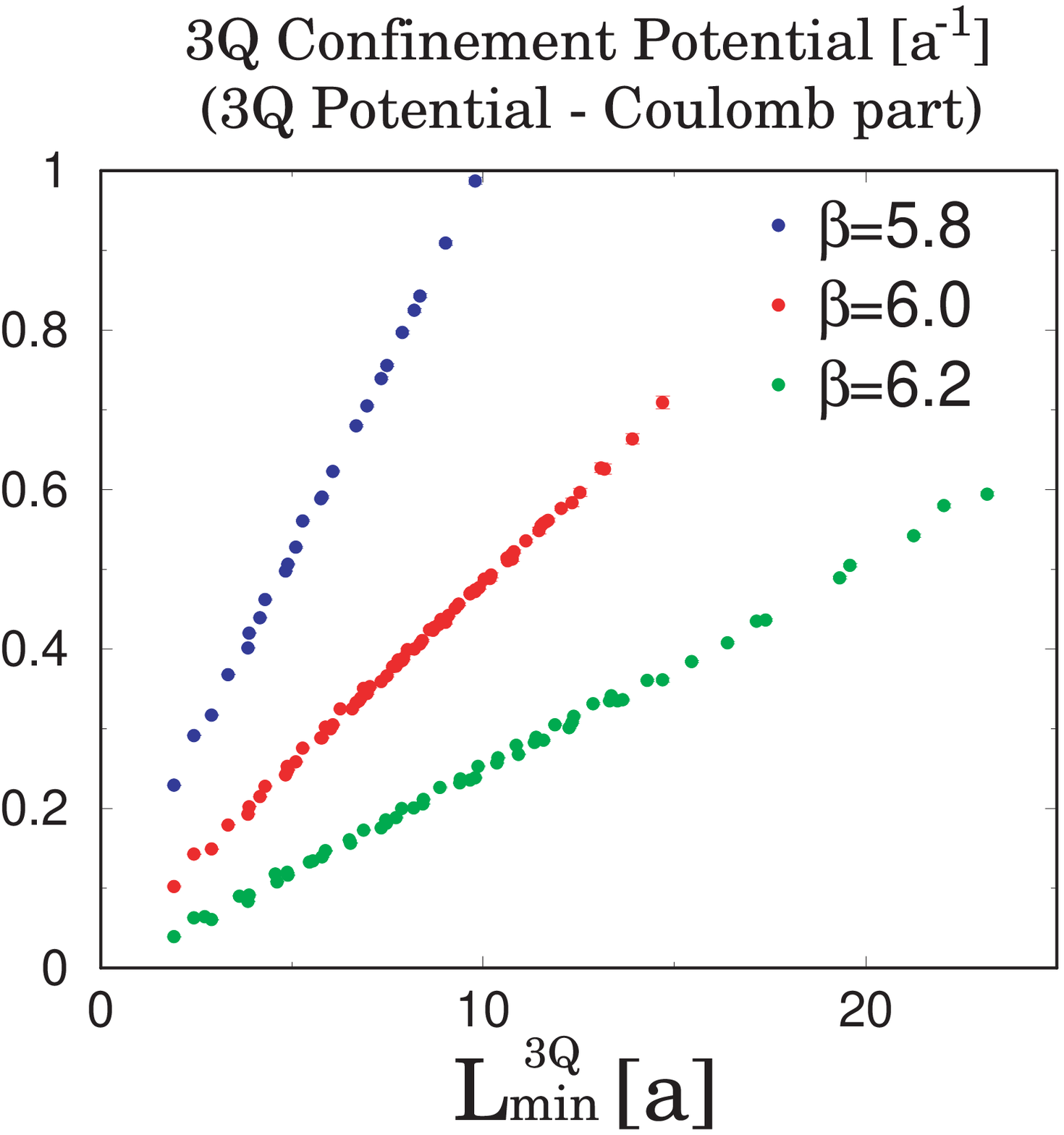}
\end{minipage}
\begin{minipage}{40mm}
\centering
\includegraphics[width=4.1cm,clip]{Fig2b.ps}
\end{minipage}
\begin{minipage}{47mm}
\centering
\includegraphics[width=5.2cm,clip]{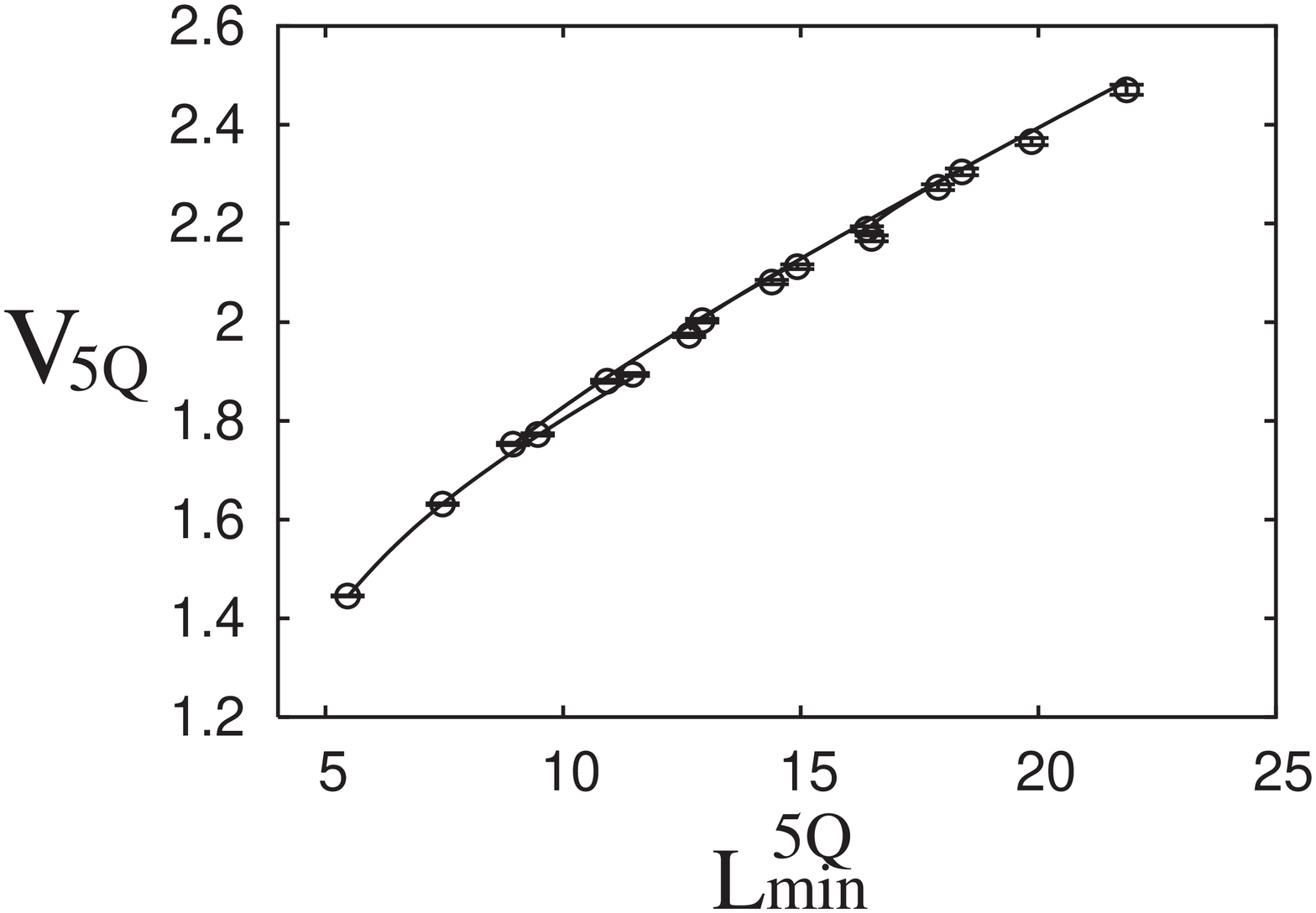}
\end{minipage}
\caption{
The lattice QCD results for the 3Q, 4Q and 5Q potentials.
(a) the 3Q confinement potential (the Coulomb-subtracted 3Q potential)
$V_{\rm 3Q}^{\rm conf}$ v.s. $L_{\rm min}^{\rm 3Q}$,
(b) ``flip-flop" in $V_{\rm 4Q}$, which takes 
the smaller energy of the multi-Y 4Q state or the two-meson state, 
(c) $V_{\rm 5Q}$ v.s. $L_{\rm min}^{\rm 5Q}$. (For details, see Ref.\cite{OST05}.)
}
\vspace{-0.5cm}
\end{figure}

We study the 4Q potential $V_{\rm 4Q}$ in lattice QCD 
for QQ-${\rm \bar{Q}\bar{Q}}$ configurations \cite{OST05}. 
When QQ and $\rm \bar Q \bar Q$ are well separated,  
$V_{\rm 4Q}$ is well described 
by the OGE Coulomb plus multi-Y Ansatz with a single connected flux-tube. 
When the nearest Q$\rm \bar{Q}$ pair is spatially close, 
$V_{\rm 4Q}$ is well described 
by the sum of two Q$\bar {\rm Q}$ potentials, 
which indicates a ``two-meson" state. 
Thus, $V_{\rm 4Q}$ is found to take 
the smaller energy of two states.
In other words, we observe a clear lattice QCD evidence 
of ``flip-flop", i.e., flux-tube recombination between the two states.
As for 5Q systems, 
we study QQ-$\rm \bar{Q}$-QQ systems and find 
that the 5Q potential $V_{\rm 5Q}$ is well described by the OGE Coulomb 
plus multi-Y Ansatz\cite{OST05}. 
Thus, from a series of our lattice QCD studies, 
the inter-quark potential is clarified to consist of 
the OGE Coulomb part  
and the flux-tube-type linear confinement part 
in both mesons, baryons and multi-quark hadrons \cite{TS,OST05}. 

\begin{figure}[h]
\begin{minipage}{60mm}
\centering
\includegraphics[height=2.2cm,clip]{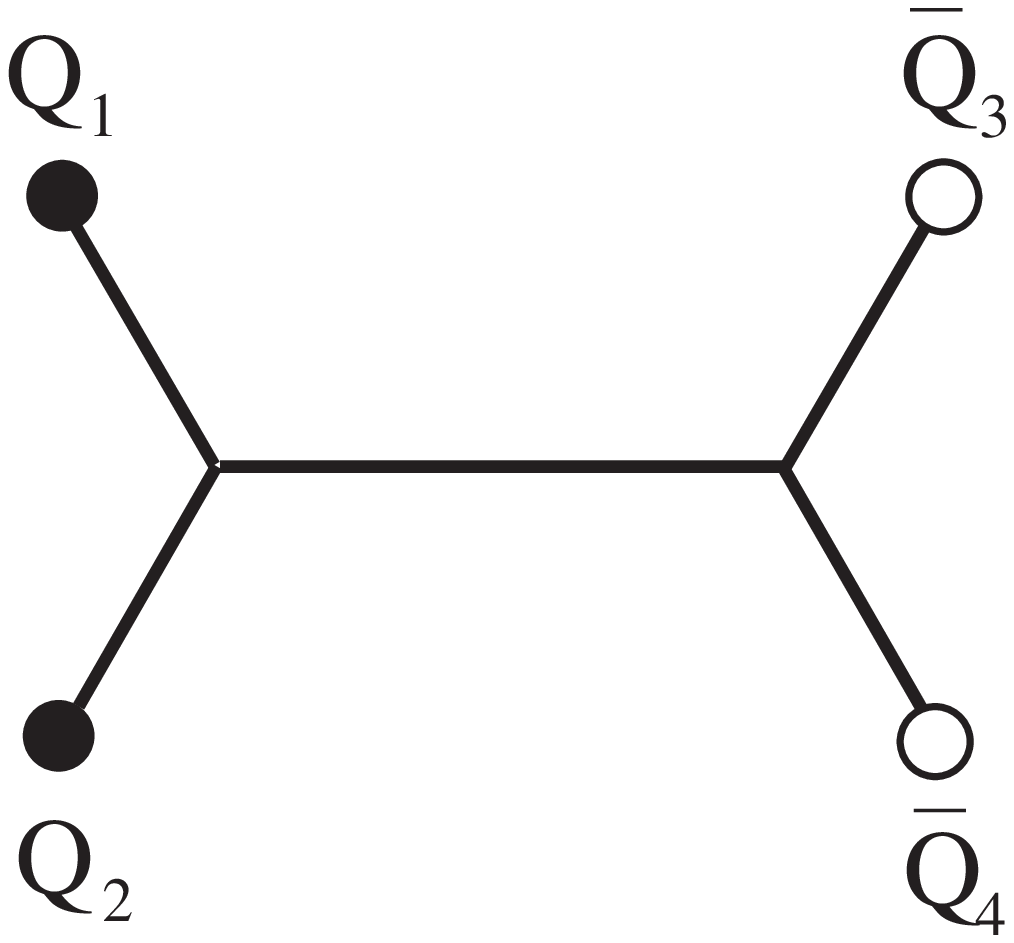}
\end{minipage}
\begin{minipage}{60mm}
\centering
\includegraphics[height=2.5cm,clip]{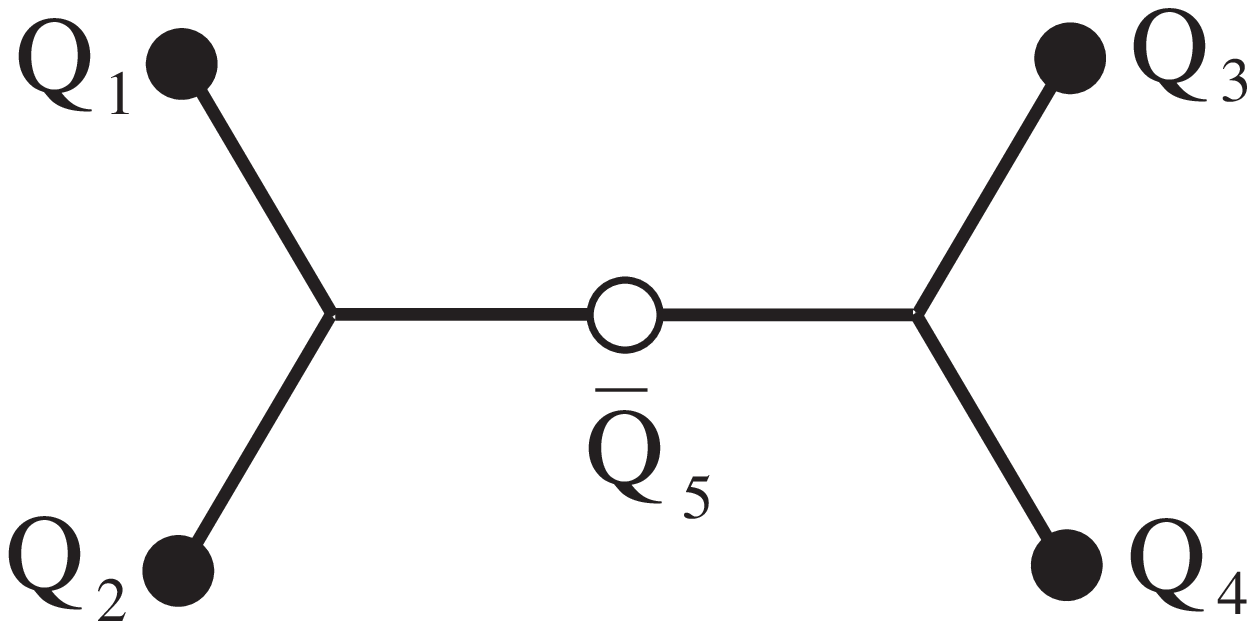}
\end{minipage}
\caption{
The flux-tube formation in multi-quarks 
indicated by our lattice QCD results.
}
\end{figure}

From the comparison among 
the $\rm Q\bar Q$, 3Q, 4Q and 5Q potentials in lattice QCD,
we find the universality of the string tension $\sigma$ \cite{TS,OST05} as 
$\sigma_{\rm Q\bar{\rm Q}}\simeq \sigma_{\rm 3Q} \simeq \sigma_{\rm 4Q} 
\simeq \sigma_{\rm 5Q}$, 
and the OGE Coulomb coefficient, i.e., $g \simeq 1.6$.
Here, the OGE Coulomb term is considered to originate 
from the OGE process, which plays the dominant role at short distances, 
where perturbative QCD is applicable.
The flux-tube-type linear confinement  
physically indicates the flux-tube picture, 
where quarks and antiquarks are linked 
by the one-dimensional squeezed color-electric flux-tube  
with the string tension $\sigma \simeq$ 0.89GeV/fm. 

To conclude, the inter-quark interaction would 
be generally described by the sum of the OGE part  
and the flux-tube-type linear confinement part 
with the universal string tension $\sigma$.
Thus, based on the lattice QCD results, we propose 
the proper quark-model Hamiltonian $\hat H$ for multi-quark hadrons as 
\begin{eqnarray}
\hat H=\sum_{i} \sqrt{\hat {\bf p}_i^2+M_i^2}+
\sum_{i<j} V_{\rm OGE}({\bf r}_i-{\bf r}_j)
+\sigma L_{\rm min},
\end{eqnarray}
where $V_{\rm OGE}$ denotes the OGE potential and 
$M_i$ the constituent quark mass.

To summarize, we have performed the multi-quark potential in lattice QCD 
to clarify the inter-quark interaction in multi-quark systems. 
We have found that the 3Q potential is well described by the Y-Ansatz. 
For 4Q and 5Q systems,
$V_{\rm 4Q}$ and $V_{\rm 5Q}$ are well described 
by the OGE Coulomb plus multi-Y Ansatz,
which indicates the flux-tube picture even for multi-quarks. 
We have observed a clear lattice QCD evidence of ``flip-flop''. 

As a successive work, we are now investigating the nuclear force in lattice QCD \cite{DT05}. 

The lattice calculations have been done on NEC-SX5(Osaka U.) and SR8000(KEK).


\bibliographystyle{aipproc}   

\end{document}